\documentclass[review]{elsarticle}
\usepackage{amssymb}

\journal{Physica A: Statistical Mechanics and its Applications}

\begin{document}

\begin{frontmatter}

\title{Three-Particle Correlations \\ in Liquid and Amorphous Aluminium}

\author[adr1,adr2]{Bulat N.~Galimzyanov\corref{cor1}}
\ead{bulatgnmail@gmail.com}\cortext[cor1]{Corresponding author}
\author[adr1,adr2]{Anatolii V.~Mokshin}
\ead{anatolii.mokshin@mail.ru}

\address[adr1]{Institute of Physics, Kazan
Federal University, 420008 Kazan, Russia}
\address[adr2]{Landau Institute for Theoretical Physics, Russian Academy of
Sciences, 142432 Chernogolovka, Russia}

\begin{abstract}
Analysis of three-particle correlations is performed on the basis of
simulation data of atomic dynamics in liquid and amorphous
aluminium. A three-particle correlation function is introduced to
characterize the relative positions of various three particles
--- the so-called triplets. Various configurations of triplets are found by calculation of pair and
three-particle correlation functions. It was found that in the case
of liquid aluminium with temperatures $1000\,$K, $1500\,$K, and
$2000\,$K the three-particle correlations are more pronounced within
the spatial scales, comparable with a size of the second
coordination sphere. In the case of amorphous aluminium with
temperatures $50\,$K, $100\,$K, and $150\,$K these correlations in
the mutual arrangement of three particles are manifested up to
spatial scales, which are comparable with a size of the third
coordination sphere. Temporal evolution of three-particle
correlations is analyzed by using a time-dependent three-particle
correlation function, for which an integro-differential equation of
type of the generalized Langevin equation is output with help of
projection operators technique. A solution of this equation by means
of mode-coupling theory is compared with our simulation results. It
was found that this solution correctly reproduces the behavior of
the time-dependent three-particle correlation functions for liquid
and amorphous aluminium.
\end{abstract}


\end{frontmatter}

\section{Introduction}

At present time the main attention is paid to study the structure of
condensed systems, which are in equilibrium or in metastable
states~\cite{Kashchiev_2000,March_Tosi_1991,
Steinhardt_Ronchetti_1983,Zahn_Maret_2003}. Often, the traditional
experimental methods of structural analysis do not allow to
correctly identify the presence of some structures in bulk systems
due to their small sizes, either low concentrations in the system,
or due to relatively short lifetimes. Usually, information about the
structure of condensed systems is extracted by using microscopic
methods or by methods of X-ray and neutron diffraction. Here, the
static structure factor determined from the experimental data is
critical quantity, which is related with the pair distribution
function, $g(r)$~\cite{March_Tosi_1991,Zahn_Maret_2003}. At the same
time, the pair distribution function $g(r)$ can be determined on the
basis of atomic/molecular dynamics simulations, and then it can be
compared with experimental data.

In addition, the three-particle correlations has an essential impact
for different processes in condensed systems. Thus, an account for
three-particle correlations is required to explain the dynamic
heterogeneity in liquids~\cite{Hurley_1996,Vorselaars_2007}, to
describe of transport properties in chemical
reactions~\cite{Zahn_Maret_2003,Lazaridis_2000}, to study the structural
heterogeneity of materials at mechanical
deformations~\cite{Wang_Dhont_2002, Mokshin_Galimzyanov_2013}, to
detect the nuclei of on ordered phase (i.e. crystalline,
quasicrystalline)~\cite{Dzugutov_1993,Doye_Wales_2001}, to describe
the amorphization of liquids at rapid
cooling~\cite{March_Tosi_1991,Zahn_Maret_2003,Tokuyama_2007}. Direct evaluation of three-particle correlations by means
of experimental measurements is extremely difficult
problem~\cite{Zahn_Maret_2003,Vaulina_Petrov_Fortov_2004,Ma_Zuo_2007}.
Here, the special methods must be adopted to extract such
information~\cite{Zahn_Maret_2003,Egelstaff_Page_1969, Montfrooij_Graaf_1991}. On the other hand, detailed information about the
three-particle correlations can be obtained on the basis of
atomic/molecular dynamics simulations data. Note that early studies
of the three-particle correlations were focused mainly on simple
liquids such as the Lennard-Jones fluid, the hard spheres system,
and the colloidal systems~\cite{Zahn_Maret_2003,Alder_1964,
Gupta_1982,McNeil_1983,Attard_1992}. In some recent
studies, on the basis of data of the molecular dynamics simulation
the three-particle correlations were estimated in such systems as
carbon nanotubes, electrolytes, metallic melts and
alloys~\cite{March_Tosi_1991,Gaskell_1988,Deilmann_2016}. In these
studies, the information about three-particle correlations is
usually extracted from the time evolution of two and more parameters
characterized the positions and trajectories of the particles
relative each other.

In the present work, the original method of three-particle
structural analysis and evaluation of time-dependent three-particle
correlations is proposed, where the arbitrary trajectories of motion
of the various three particles (that will be denoted as
triplets) are considered. The method allows one to identify the
presence of ordered crystalline and ``stable'' disordered
structures, which are difficult to be detected by conventional
methods of structural analysis (for example, such as the Voronoi's
tessellation method~\cite{Medvedev_2000}, the Delaunay's
triangulation method~\cite{Schachter_1980}, the bond-orientational
order parameters~\cite{Steinhardt_Ronchetti_1983}). The
applicability of this method will be demonstrated for the case of
the liquid and amorphous aluminium.

\section{Simulation Details}\label{s2}

We performed the atomic dynamics simulation of liquid and amorphous
aluminium. The system contains $N=864$ atoms, located into a cubic
simulation cell with periodic boundary conditions in all directions.
The interatomic forces are calculated through
EAM-potential~\cite{Ercolessi_1994,Winey_2009}. The velocities and
coordinates of atoms are determined through the integrating Newton's
equations of motion by using Velocity-Verlet algorithm with the
time-step $\Delta t=1\,$fs.

Initially, a crystalline sample with \emph{fcc} lattice and numerical
density $\rho=1.23\,\sigma^{-3}$ (or mass density
$2300\,$kg/m$^{3}$) was prepared, where $\sigma=2.86\,$\AA~is the effective
diameter of the aluminium atom. Further, the system was
melted to the temperatures $T=1000\,$K, $1500\,$K, and $2000\,$K.
The amorphous samples were generated through the fast cooling with
the rate $10^{12}\,$K/c of a melt at the temperature $2000\,$K to
the temperatures $T=50\,$K, $100\,$K, and $150\,$K (the melting
temperature is $T_{m}\simeq934\,$K). The simulations were performed
in \emph{NpT}-ensemble at constant pressure $p=1\,$atm.

\section{Methods}\label{s3}
\subsection{Three-particle correlation function}\label{s3_1}

\begin{figure}[h!]
\centering
\includegraphics[width=90mm]{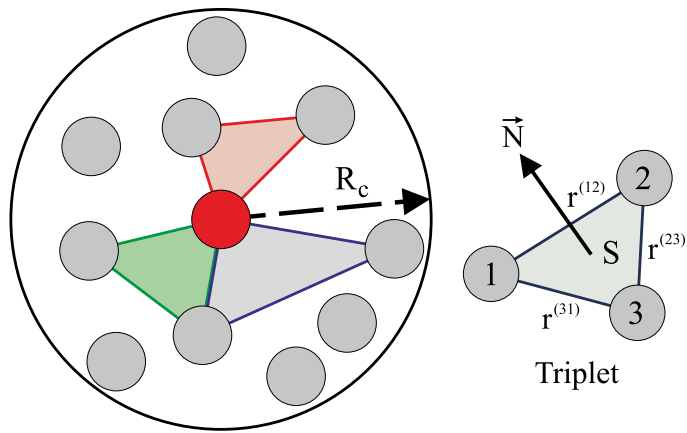}
\caption{(Color online) Schematic illustration demonstrated the
triplets with single and general vertices as well as single triangle with a
some area $S$, where the vertices with labels $1$, $2$, and $3$ are
presented.}\label{fig_1}
\end{figure}

Let us consider a system, consisting of $N$ classical particles with
same masses $m$, which are located into the simulation cubic cell
with a volume $V$. From the geometric point of view, locations of
any three particles generate a triangle (i.e. triplet). This triplet
is characterized by the area, $S$. Then, the
area of $i$th triplet, $S_{i}$, at time $t$ (here $i\in\{1,2,
...,N_{T}\}$, $N_{T}$ is the number of all possible triplets in the
system) is defined by
\begin{equation}
S_{i}(t)=\left\{l_{i}(t)\cdot[l_{i}(t)-r_{i}^{(12)}(t)]\cdot[l_{i}(t)-r_{i}^{(23)}(t)]\cdot[l_{i}(t)-r_{i}^{(31)}(t)]
\right\}^{1/2}.\label{eq_1}
\end{equation}
Here, $l_{i}(t)=[r_{i}^{(12)}(t)+r_{i}^{(23)}(t)+r_{i}^{(31)}(t)]/2$
is the semiperimeter of $i$th triplet; $r_{i}^{(12)}$,
$r_{i}^{(23)}$, and $r_{i}^{(31)}$ are the distances between the
vertices of $i$th triplet with conditional labels $1$, $2$, and $3$. It follows from Eq.(\ref{eq_1}) that the area
of $i$th triplet, $S_{i}$, takes positive values and values closed
to zero. Different triplets can be correlated to the same values
$S_{i}$, and the triplets are independent from each other (i.e. have
no common vertices) or interconnected (i.e. one or two vertices are
mutual). To estimate the quantity $S_{i}$ one vertex of triplet is
considered as central (i.e. fixed), relative to which the positions
of the other vertices must be defined [see Fig.\ref{fig_1}].

To determine the probability of emergence of the triplets with the
area $S$, the three-particle correlation function is introduced
\begin{equation}
g(S)=\frac{1}{N_{T}}
\sum_{i=1}^{N_{T}}\delta(S-S_{i}).\label{eq_3}
\end{equation}
Here, the number of the all triplets in the system with $N$ particles is
defined by
\begin{equation}
N_{T}=\frac{N(N-1)(N-2)}{6}.\label{eq_2}
\end{equation}
It follows from Eq.(\ref{eq_2}) that for a system with $N=500$
particles the number of triplets $N_{T}$ is more than
$20\,000\,000$. It demonstrates that treatment of simulation results
requires significant computing resources. Therefore, at realization
of the three-particle structural analysis we can restrict our
attention to consider a spherical region with a fixed radius $R_{c}$
in the center of which the main (i.e. fixed) particle of triplet is
located. The optimal value of radius $R_{c}$ corresponds to the
distance at which the pair correlation function $g(r)$ of considered
system ceases to oscillate.

\subsection{Dynamics of three-particle correlations}\label{s3_2}

In a system with $N$ particles the coordinates $\vec{q}$ and
impulses $\vec{p}$ form a $6N$-dimensional phase space. The time
evolution of the system will be defined by the Hamiltonian
$H(\vec{q},\vec{p})$, which can be written through the canonical
Liouville equation of motion as follows
\cite{Mokshin_Chvanova_Khrm_2012}:
\begin{equation}
\frac{dA(t)}{dt}=\{H(\vec{q},\vec{p}),\,A(t)\}=\sum_{i=1}^{N}\left(\frac{\partial
H(\vec{q},\vec{p})}{\partial p_{i}}\frac{\partial A(t)}{\partial
q_{i}}-\frac{\partial H(\vec{q},\vec{p})}{\partial
q_{i}}\frac{\partial A(t)}{\partial p_{i}}\right)\label{eq_5}
\end{equation}
or
\begin{equation}
\frac{dA(t)}{dt}=iLA(t),\label{eq_6}
\end{equation}
where $L$ is the Liouville operator, $\{...\}$ is the Poisson
brackets, $A$ is the dynamic variable that obtained from results of
simulation. By means of the technique of Zwanzig-Mori's projection
operators
\begin{equation}
\Pi_{0}=\frac{A_{0}(0)\big\rangle\big\langle
A_{0}^{*}(0)}{\left\langle\left|A_{0}(0)\right|^{2}\right\rangle},\,\,P_{0}=1-\Pi_{0},\label{eq_7}
\end{equation}
one obtains from Eq.(\ref{eq_6}) the following non-Markovian
equation
\cite{Mokshin_Chvanova_Khrm_2012,Zwanzig_2001,Yulmetyev_2005,Yulmetyev_2009}:
\begin{equation}
\frac{dF(t)}{dt}=-\Omega_{1}^{2}\int_{0}^{t}M_{1}(\tau)F(t-\tau)d\tau,\label{eq_8}
\end{equation}
where
\begin{equation}
F(t)=\frac{\langle
A_{0}^{*}(0)A_{0}(t)\rangle}{\langle|A_{0}(0)|^{2}\rangle}\label{eq_9}
\end{equation}
is the time correlation function;
\begin{equation}
M_{1}(\tau)=\frac{\langle
A_{1}^{*}(0)e^{iL_{22}^{0}\tau}A_{1}(0)\rangle}{\langle|A_{1}(t)|^{2}\rangle}\label{eq_11}
\end{equation}
is the first order memory function;
\begin{equation}
\Omega_{1}^{2}=\frac{\langle|A_{1}(0)|^{2}\rangle}{\langle|A_{0}(0)|^{2}\rangle}\label{eq_12}
\end{equation}
is the first-order frequency parameter, and
\begin{equation}
A_{1}(t)=iLA_{0}(t)\label{eq_13}
\end{equation}
is the next dynamics variable. By considering that the Liouville
equation for dynamic variable $A_{1}(t)$
\begin{equation}
\frac{dA_{1}(t)}{dt}=iL_{22}^{0}A_{1}(t),\label{eq_14}
\end{equation}
the kinetic integro-differential equation can be obtained for the
first-order memory function as follows
\cite{Mokshin_Chvanova_Khrm_2012}:
\begin{equation}
\frac{dM_{1}(t)}{dt}=-\Omega_{2}^{2}\int_{0}^{t}M_{2}(\tau)M_{1}(t-\tau)d\tau.\label{eq_15}
\end{equation}
Similarly, a chain of equations can be constructed as
\begin{equation}
\frac{dM_{n-1}(t)}{dt}=-\Omega_{n}^{2}\int_{0}^{t}M_{n}(\tau)M_{n-1}(t-\tau)d\tau,\,n=1,\,2,\,3,...\label{eq_16}
\end{equation}
Here, the $n$th order frequency parameter will be determined as
follows \cite{Mokshin_Chvanova_Khrm_2012,Mokshin_tmf_2015}:
\begin{equation}
\Omega_{n}^{2}=\frac{\langle|A_{n}(0)|^{2}\rangle}{\langle|A_{n-1}(0)|^{2}\rangle}.\label{eq_17}
\end{equation}
Using the Laplace transformation,
$\widetilde{M}_{n}(s)=\int_{0}^{\infty}e^{-st}M_{n}(t)dt$ (where
$s=i\omega$), the chain of equations (\ref{eq_16}) can be rewritten
as a continued fraction \cite{Mokshin_Chvanova_Khrm_2012}
\begin{equation} \widetilde{F}(s)=\frac{1}{\displaystyle s+\frac{\Omega_{1}^{2}}{\displaystyle s+\frac{\Omega_{2}^{2}}{\displaystyle s+\ldots}}}.\label{eq_18}
\end{equation}

From Eq.(\ref{eq_16}) one obtains
\cite{Mokshin_Chvanova_Khrm_2012,Mokshin_tmf_2015}
\begin{equation}
\ddot{F}(k,t)+\Omega_{1}^{2}(k)F(k,t)+\Omega_{2}^{2}(k)\int_{0}^{t}M_{2}(k,t-\tau)\dot{F}(k,\tau)d\tau=0,\label{eq_19}
\end{equation}
where $k=|\vec{k}|$ is the wave number
\cite{March_Tosi_1991,Zwanzig_2001}. By using the key condition of
the mode-coupling theory
\cite{Mokshin_Chvanova_Khrm_2012,Mokshin_tmf_2015}
\begin{equation}
\Omega_{2}^{2}(k)M_{2}(k,t)=\varphi(k)\delta(t)+\Omega_{1}^{2}(k)\left[\upsilon_{1}F(k,t)+\upsilon_{2}F(k,t)^{p}\right],\label{eq_20}
\end{equation}
the Eq.(\ref{eq_19}) can be rewritten in the form
\begin{eqnarray}
\ddot{F}(k,t)+\Omega_{1}^{2}(k)F(k,t)+\varphi(k)\dot{F}(k,t)\delta(t)+ \nonumber \\ +\Omega_{1}^{2}(k)\int_{0}^{t}\left[\upsilon_{1}F(k,t-\tau)+\upsilon_{2}F(k,t-\tau)^{p}\right]\dot{F}(k,\tau)d\tau=0.\label{eq_21}
\end{eqnarray}
Here, $\Omega_{1}(k)$ and $\varphi(k)$ are the frequency parameters,
$\delta(t)$ is the Dirac's delta function, $\upsilon_{1}\geq0$,
$\upsilon_{2}\geq0$ ($\upsilon_{1}+\upsilon_{2}\neq0$) are the
weight of the corresponding contributions, the parameter $p>1$ can
be fractional. The exact solution of Eq.(\ref{eq_21}) will be
defined through the frequency parameters $\Omega_{1}(k)$,
$\varphi(k)$ as well as through the characteristics $\upsilon_{1}$,
$\upsilon_{2}$, and $p$ \cite{Khusnutdinov_Mokshin_2010}.

A numerical solution of integro-differential Eq.(\ref{eq_21})
can be found from~\cite{Khusnutdinov_Mokshin_2010}:
\begin{equation}\label{eqS_1}
z_{n}+\Omega_{1}^{2}x_{n}+\Omega_{1}^{2}\tau\sum_{i=0}^{n}\left[\upsilon_{1}x_{i}+\upsilon_{2}x_{i}^{p}\right]z_{n-i}=0,
\end{equation}
\begin{equation}\label{eqS_2}
y_{n+1}=y_{n}+\tau z_{n},
\end{equation}
\begin{equation}\label{eqS_3}
x_{n+1}=x_{n}+\tau y_{n+1}.
\end{equation}
Let us take the quantity
\begin{equation}
\vec{s}_{j}(t)=\frac{\mu_{j}S_{j}(t)\vec{N}_{j}}{\sigma},\,\,j=1,2,...,N_{T} \label{eqS_4}
\end{equation}
as a dynamical variable. If the value $A_{0}=\vec{s}_{j}(t)$ is
taken as initial dynamical variable [see Eq.(\ref{eq_7})], then the
time correlation function will be defined as
\begin{equation}
F_{T}(k,t)=\frac{1}{N_{T}}\sum_{j=1}^{N_{T}}\exp\left[-i\vec{k}(\vec{s}_{j}(t)-\vec{s}_{j}(0))\right],\label{eqS_5} 
\end{equation}
and $x_{n}=F_{T}(k,t)$, $y_{n}=\dot{F}_{T}(k,t)$,
$z_{n}=\ddot{F}_{T}(k,t)$, where $\tau=0.01\,$fs, while initial
conditions in Eq.(\ref{eqS_1}) are $F_{T}(k,t=0)=1$ and $\dot{F}_{T}(k,t=0)=0$
\cite{Khusnutdinov_Mokshin_2010}. Here, $S_{j}(t)$ is the area of $j$th
triplet at time $t$,
$\vec{N}_{j}=n_{j1}\vec{e_{x}}+n_{j2}\vec{e_{y}}+n_{j3}\vec{e_{z}}$
is the vector of normal to the plane of this triangle [see illustration in
Fig.\ref{fig_1}],
$\mu_{j}=\pm[n_{j1}^{2}+n_{j2}^{2}+n_{j3}^{2}]^{-1/2}$ is the
normalization constant determined by a sign of parameter $n_{j4}$
from equation of the plane $n_{j1}x+n_{j2}y+n_{j3}z+n_{j4}=0$, which
connects of vertices of $j$th triplet. At $n_{j4}>0$ we have a value
$\mu_{j}<0$, otherwise we have a value $\mu_{j}>0$.

\section{Results and Discussions}\label{s4}
\subsection{Behavior of single triplet}\label{s4_2}

\begin{figure}[h!]
\centering
\includegraphics[width=90mm]{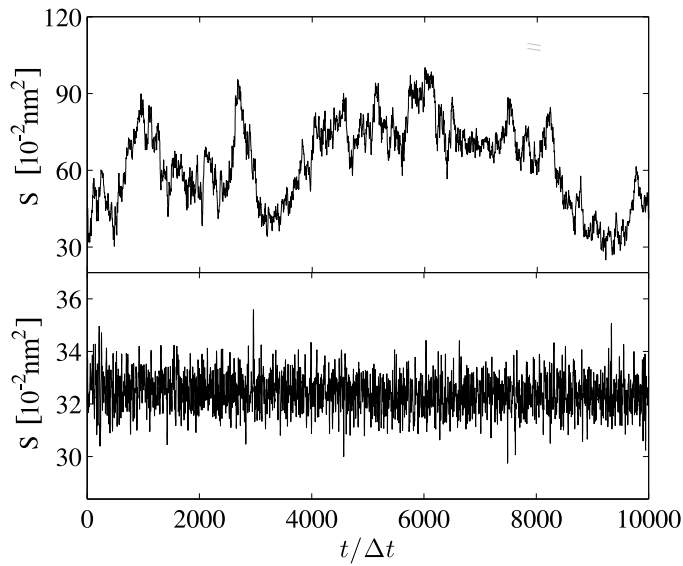}
\caption{Time-dependent area of a triplet in atomic dynamics of
aluminium. Top panel: liquid aluminium at the temperature $1000$\,K.
Bottom panel: amorphous aluminium at the temperature
$100$\,K.}\label{fig_2}
\end{figure}

The time dependent area of a triplet is evaluated at different
temperatures. Thus, in Fig.\ref{fig_2} (top panel) it is shown the
time-dependent area of a triplet $S(t)$ in liquid aluminium at
temperature $1000\,$K. It can be seen that the quantity $S(t)$ takes
the values in the range $0.25\,$nm$^{2}<S<1\,$nm$^{2}$. The
evolution of $S(t)$ in the case of liquid aluminium is characterized
by the high-frequency fluctuations of a relatively small amplitude,
which are originated due to collective motion of atoms. The
low-frequency fluctuations with a relatively large amplitude define
a main trend of $S(t)$, and they defined by transition to the
diffusive motion of particles, when an atom or several atoms leave a
nearest environment.

Fig.\ref{fig_2} (bottom panel) represents the time-dependent area
of a triplet in amorphous aluminium at temperature $T=100\,$K.
Contrasted to the liquid, the diffusive regime of $S(t)$ is not
observed, that is due to extremely low mobility of atoms. The
fluctuations with relatively small amplitude are seen; wherein the
area $S(t)$ takes the values within a narrow range
$0.29\,$nm$^{2}<S<0.36\,$nm$^{2}$. These fluctuations is caused by
vibrations of atoms in a surrounded of ``neighbors''.

\subsection{Features of three-particle correlations}\label{s4_2}

The distribution function $g(S)$ was computed by Eq.(\ref{eq_3}) for the liquid and
amorphous systems, and those triplets were considered, only which
located within a sphere of the radius $R_{c}=3\,\sigma$. Also, the
pair distribution function was determined as follows
\cite{March_Tosi_1991,Poole_1998,Khusnutdinoff_2012}:
\begin{equation}
g(r)=\frac{V}{4\pi r^{2}N}\sum_{i=1}^{N}\left\langle\frac{\Delta n_{i}(r)}{\Delta r}\right\rangle, \label{eq_rdf}
\end{equation}
where $\Delta n_{i}(r)$ is the probability to find the pair of the
atoms separated by the distance $r$. For convenience, the distances between the atoms we will measure in the units of $\sigma$ and the area of triplet in the units of $\sigma^{2}$ (here, $\sigma=2.86\,$\AA).

\begin{figure}[h!]
\centering
\includegraphics[width=130mm]{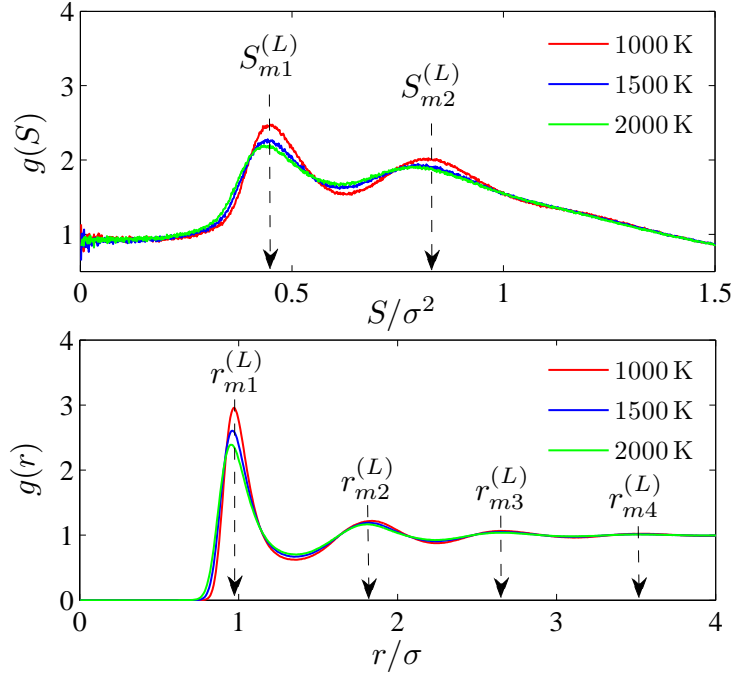}
\caption{(Color online) Three-particle correlation function $g(S)$
(top panel) and pair distribution function $g(r)$ (bottom panel),
obtained for liquid aluminium at temperatures $1000$\,K, $1500$\,K,
and $2000$\,K. The positions of main maxima of the functions $g(S)$
and $g(r)$ are indicated by arrows.}\label{fig_3}
\end{figure}

\begin{figure}[h!]
\centering
\includegraphics[width=100mm]{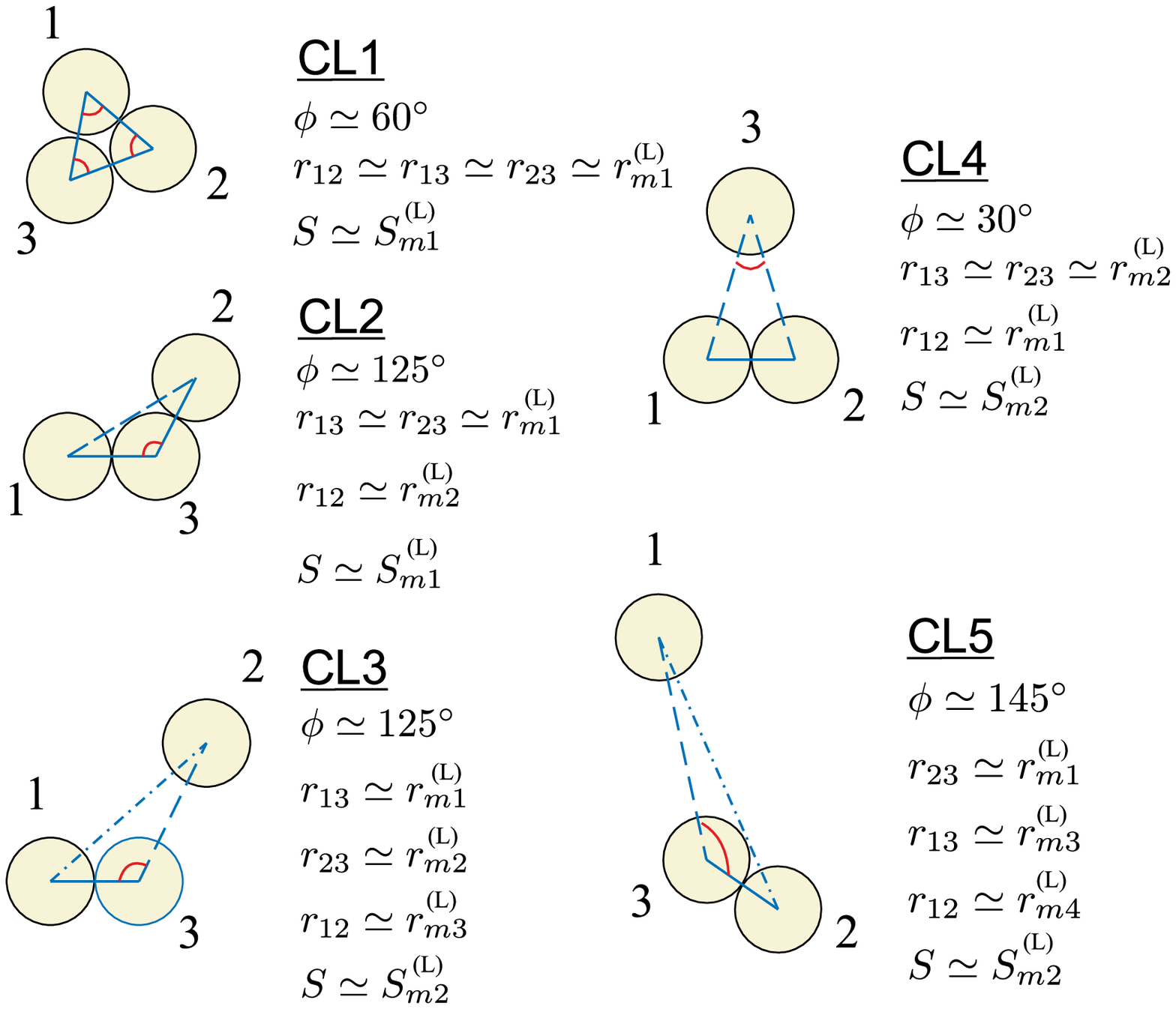}
\caption{(Color online) Triplets which appear in liquid aluminium
and correspond to the main maxima of the functions $g(S)$ and
$g(r)$. The circles mark the atoms. The lines mark the distances
$r_{12}$, $r_{23}$, and $r_{13}$ between the atoms. Here, $\phi$ is
the angle between two straight lines, which connect the vertices of
the triangle.}\label{fig_4}
\end{figure}

Fig.\ref{fig_3} shows the curves $g(S)$ and $g(r)$ for liquid
aluminium at temperatures $1000\,$K, $1500\,$K, and $2000\,$K. As
seen the pair distribution function $g(r)$ of liquid aluminium has
oscillations and contains the maxima at distances $r=r_{m1}^{(L)}$,
$r_{m2}^{(L)}$, $r_{m3}^{(L)}$, and $r_{m4}^{(L)}$, which characterize the correlation
lengths. The estimated inter-particle distances $r=r_{m1}^{(L)}$,
$r_{m2}^{(L)}$, $r_{m3}^{(L)}$, and $r_{m4}^{(L)}$ are given in Table \ref{table_1}. The undistinguished maximum at
$r\simeq r_{m4}^{(L)}$ is indication that the pair correlations are practically
absent at large distances. For the reason that both the functions
$g(S)$ and $g(r)$ correspond to the same system, it is quite
reasonable to assume that the maxima in function $g(S)$, located at
$S=S_{m1}^{(L)}$ and $S_{m2}^{(L)}$, are associated with the triplets, in which
the distances between pairs of atoms correspond to correlation
lengths $r_{m1}^{(L)}$, $r_{m2}^{(L)}$, $r_{m3}^{(L)}$, and $r_{m4}^{(L)}$. Then, the
different possible configurations of three atoms can be selected,
where the atoms that form these triplets are remote to the distances
$r=r_{m1}^{(L)}$, $r_{m2}^{(L)}$, $r_{m3}^{(L)}$, and $r_{m4}^{(L)}$. As a result of such
selection, five different widespread configurations appear in liquid
aluminium, which are depicted in Fig.\ref{fig_4} with labels CL1, CL2,..., CL5.

From analysis of simulation data, it follows that the triplets with
configurations CL1 and CL2, which correspond to the maximum at
$S_{m1}^{(L)}$ in the function $g(S)$ that also give impact to the main
maxima of the function $g(r)$ at $r_{m1}^{(L)}$ and $r_{m2}^{(L)}$. In this
case, a mutual arrangement of three atoms covers a spatial scale,
which are comparable with size of the first and second coordination
spheres. Also, the presence of triplets with configurations CL1 and CL2 is observed in other model liquids~\cite{Zahn_Maret_2003}.  So, Zahn et al. observed the triplets with inter-particle distances $r\simeq1.0\,\sigma$ and $r\simeq1.9\,\sigma$ (i.e. with configurations similar to CL1 and CL2, where $r=r_{m1}^{(L)}\simeq0.96\,\sigma$ and $r_{m2}^{(L)}\simeq1.82\,\sigma$) in two-dimensional colloidal liquid, where three-particle correlation functions were calculated from particle configurations (see Fig.2 and Fig.4 in Ref.\cite{Zahn_Maret_2003}). The triplets CL3, CL4, and CL5 that
correspond to the maximum of function $g(S)$ at $S_{m2}^{(L)}$, are
usually involved to formation of maxima of the function $g(r)$ at
$r_{m3}^{(L)}$ and $r_{m4}^{(L)}$. Remarkable that the structure of the system
can be restored by connecting the different configurations CL1, CL2,
CL3, CL4, and CL5.

\begin{figure}[h!]
\centering
\includegraphics[width=120mm]{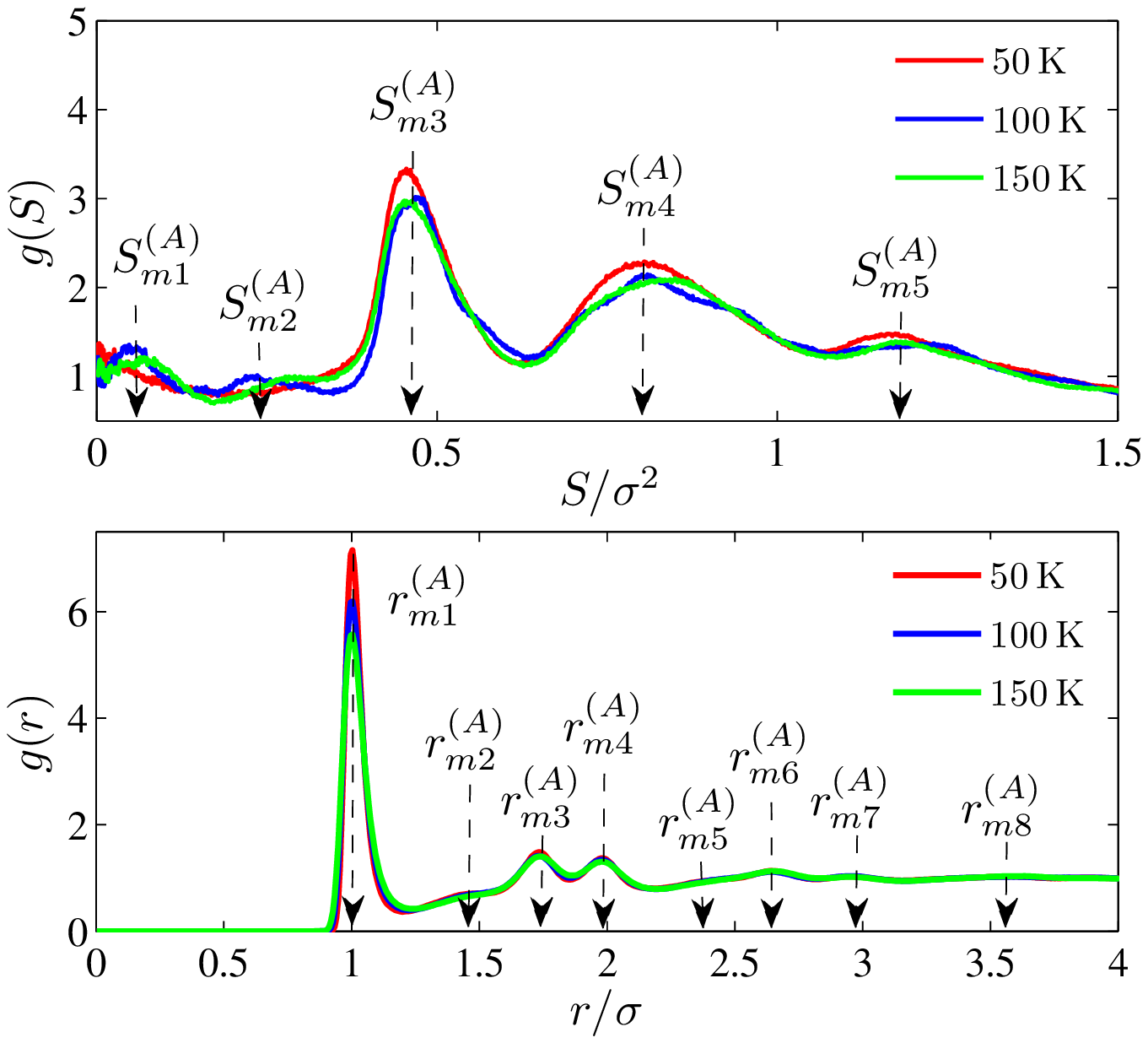}
\caption{(Color online) Three-particle correlation function $g(S)$
(top panel) and pair distribution function $g(r)$ (bottom panel),
obtained for amorphous aluminium at temperatures $50\,$K, $100\,$K,
and $150\,$K.}\label{fig_5}
\end{figure}

Fig.\ref{fig_5} represents the functions $g(S)$ and $g(r)$ for
amorphous system at temperatures $50\,$K, $100\,$K, and $150\,$K.
Five pronounced maxima are detected for $g(S)$, and eight maxima are
most pronounced for the function $g(r)$. As seen from
Fig.\ref{fig_5} (top panel), the function $g(S)$ contain the maxima
at $S=S_{m1}^{(A)}$, $S_{m2}^{(A)}$, $S_{m5}^{(A)}$, as $S_{m3}^{(A)}$ and $S_{m4}^{(A)}$, which are
not observed for liquid system. The estimated inter-particle
distances, which correspond to positions of maxima of the functions
$g(S)$ and $g(r)$, are given in Table \ref{table_2}.

\begin{figure}[h!]
\centering
\includegraphics[width=120mm]{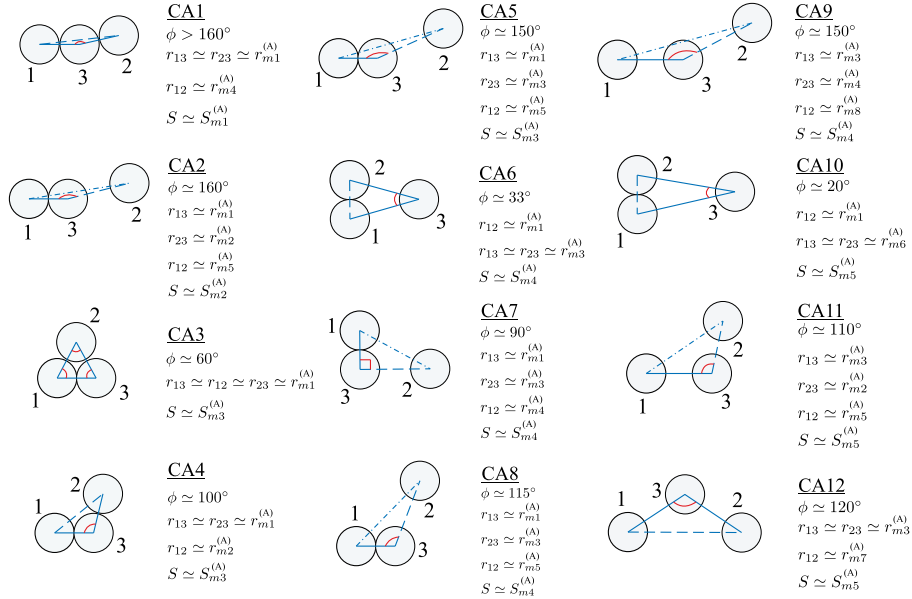}
\caption{(Color online) Configurations of triplets in the amorphous
aluminium, which involve in the formation of the main maxima of the
functions $g(S)$ and $g(r)$.}\label{fig_6}
\end{figure}

In the case of amorphous aluminium, various triplets are detected,
where twelve configurations can be chosen as widespread, which depicted in Fig.\ref{fig_6} with labels CA1, CA2,..., CA12. Wherein, the most of the configurations except of
CA1, CA3, and CA4 cover the
spatial scale, which exceeds the size of the second coordination
sphere. The presence of maxima at small areas $S$ (for example, the
maxima at $S=S_{m1}^{(A)}$ and $S_{m2}^{(A)}$), which correspond to CA1 and CA2, can be indication of structural ordering, as
well as evidence of the presence of quasi-ordered structures. At the same time, the configurations
CA1 and CA2 together with CA3 and CA4 generate the main maximum of
the function $g(r)$ at the distance $r=r_{m1}^{(A)}$. The formation of
triples CA5, CA6,..., CA9, that correspond to the maxima of function
$g(S)$ at $S=S_{m3}^{(A)}$ and $S_{m4}^{(A)}$, leads to splitting of second
maximum of the function $g(r)$. The configurations CA2, CA4, and
CA11 participate to formation of the maximum of $g(r)$ at
$r=r_{m2}^{(A)}$, and give impact to the maxima of the three-particle
correlation function $g(S)$ at $S=S_{m2}^{(A)}$, $S_{m3}^{(A)}$, and $S_{m5}^{(A)}$.
The presence of maximum in the function $g(S)$ at $S=S_{m5}^{(A)}$ is due to
the triplets CA10, CA11,..., CA12, the number of which is
negligible.

The configurations CL1, CL2,..., CL5 and CA1, CA2,..., CA12 appear
because the atoms that form these triplets are surrounded by
neighbors, which slow down their motion -- so-called cage effect.
According to effective neighborhood model proposed by Vorselaars et
al., the cage effect occurs in liquids and glasses, where the
particles are trapped in a local energy
minimum~\cite{Vorselaars_2007}. Namely, a single particle at motion jumps from one cage to another cage (i.e. from one local energy minimum to another local energy minimum)~\cite{Vorselaars_2007}. In the case of aluminium atoms with configurations, which are depicted in Fig.\ref{fig_4} and Fig.\ref{fig_6}, the jumps between different cages lead to transitions between various triplets. Usually, these transitions occur between triplets with
similar configurations, for example, between CL1 and CL2, between
CL3 and CL4 in the case of liquid aluminium as well as between CA1
and CA2, between CA3 and CA4, between CA6 and CA10, between CA7 and
CA8 in the case of amorphous aluminium. Unlike liquid, in the
amorphous system the transitions between the triplets of different
configurations occur very slowly, where the jumps of atoms between
different cages occur for a long time due to the high viscosity.

\subsection{Time-dependent three-particle correlations}\label{s4_3}

\begin{figure}[h!]
\centering
\includegraphics[width=90mm]{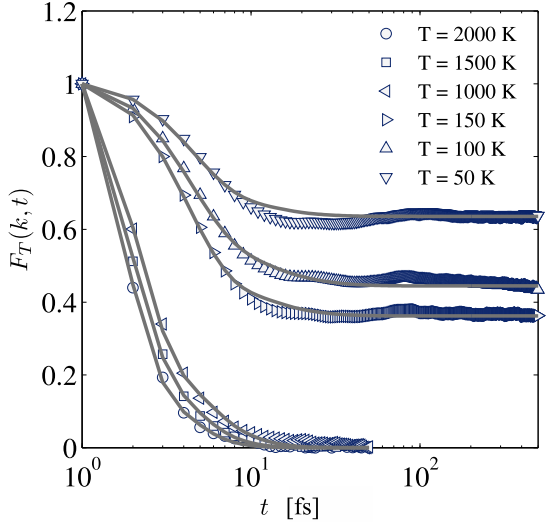}
\caption{(Color online) Three-particle correlation function
$F_{T}(k,t)$ obtained for aluminium at various temperatures and at
wave number $k=4.6$\,\AA$^{-1}$. The solid lines mark the
theoretical results; symbols depict results of
simulations.}\label{fig_7}
\end{figure}

The analysis of the time-dependent three-particle correlations is
also done by means of the time correlation function $F_{T}(k, t)$,
defined by Eq.(\ref{eqS_5}). As an example, the function $F_{T}(k,
t)$ obtained for liquid and amorphous aluminium at different
temperatures is presented in Fig.\ref{fig_7}. The calculations were
performed at fixed value of the wave number $k=4.6\,$\AA$^{- 1}$. We
note that the behavior of the function $F_{T}(k,t)$ is similar to
behavior of incoherent scattering function
$F(k,t)$~\cite{Hansen_McDonald_2006}. The function $F_{T}(k,t)$
demonstrates the fast decay to zero in the case of liquid system,
that is caused with attenuation of three-particle correlations. The
rate of attenuation is expected to be correlated with the structural
relaxation time. In the case of amorphous aluminium at temperatures
$50\,$K, $100\,$K, and $150\,$K, the function $F_{T}(k,t)$ is
characterized by more complex shape.

On the other hand, the function $F_{T}(k,t)$ was computed by Eq.(\ref{eq_21}) on the basis of simulation results at $k=4.6\,$\AA$^{-1}$ and the
temperatures $T=50\,$K, $100\,$K, $150\,$K, $1000\,$K, $1500\,$K,
and $2000\,$K. Then, the Eq.(\ref{eq_21}) was solved numerically according to
scheme (\ref{eqS_1})--(\ref{eqS_5}), while the parameters
$\upsilon_{1}$, $\upsilon_{2}$, $\Omega_{1}^{2}$, $p$ were taken as
adjustable. The values of the parameters are given in Table
\ref{table_3}. As can be seen from Fig.\ref{fig_7}, the theory with
Eq.(\ref{eq_21}) reproduces the behavior of function $F_{T}(k,t)$ at
all considered temperatures. Further, for the case of amorphous
system, the theory allows one to reproduce the emergence of plateau
in function $F_{T}(k,t)$. The frequency parameter, $\Omega_{1}^{2}$,
takes the relatively small values
$3.3\times10^{25}\,$s$^{-2}\leq\Omega_{1}^{2}\leq9.8\times10^{25}\,$s$^{-2}$
for amorphous system and large values
$64\times10^{25}\,$s$^{-2}\leq\Omega_{1}^{2}\leq130\times10^{25}\,$s$^{-2}$
for liquid system. Here, the increase of the parameter
$\Omega_{1}^{2}$ with temperature can be due to increase of
transition rate between triplets of various configurations.

\section{Conclusion}\label{s5}

In the present work, the analysis of the three-particle correlations
in many-particle systems is suggested. By atomic dynamics simulation
of liquid and amorphous aluminium, the applicability of the proposed
method of three-particle structural analysis is demonstrated to
identify the structures, which are generated by various triplets. By
applying the calculation of the pair and three-particle distribution
functions, the triplets of various configurations are found. It has
been shown that these triplets, which are formed due to particle
correlations can cover the spatial scales, which are comparable with
sizes of the second and third coordination spheres. On the other
hand, it was found that the time evolution of the three-particle
correlations in liquid and amorphous aluminium can be recovered from
transition between triplets of various configurations. Then, it is
shown that the time-dependent three-particle correlations in these
systems are reproducible by the integro-differential equation
(\ref{eq_21}). Here, an agreement between theoretical results and our atomic dynamics simulation data is observed.

\section*{Acknowledgments}
The work was supported by the grant of the President of Russian
Federation: MD-5792.2016.2. The atomic dynamics calculations
were performed on the computing cluster of Kazan Federal University
and Joint Supercomputer Center of RAS.

\newpage

\section*{Tables}

\begin{table}[h!]
\begin{center}
\caption{Inter-particle distances $r_{m1}^{(L)}$, $r_{m2}^{(L)}$, $r_{m3}^{(L)}$, and
$r_{m4}^{(L)}$ (in units $\sigma$), as well as the areas $S_{m1}^{(L)}$ and
$S_{m2}^{(L)}$ (in units $\sigma^{2}$), which correspond to positions of
the main maxima of the functions $g(r)$ and $g(S)$ in the case of
liquid aluminium. The errors in the values are estimated to be less
than $5$\%. \label{table_1}}\centerline{}
\begin{tabular}[t]{ccccccc}
\hline
$T,\,$K & $r_{m1}^{(L)}$ & $r_{m2}^{(L)}$ & $r_{m3}^{(L)}$ & $r_{m4}^{(L)}$ & $S_{m1}^{(L)}$ & $S_{m2}^{(L)}$ \\
\hline
$1000$  & $0.969$ & $1.841$ & $2.661$  & $3.535$ & $0.45$ & $0.823$ \\
$1500$  & $0.962$ & $1.822$ & $2.654$  & $3.532$ & $0.44$ & $0.803$ \\
$2000$  & $0.955$ & $1.815$ & $2.647$  & $3.528$ & $0.43$ & $0.797$ \\
\hline
\end{tabular}
\end{center}
\end{table}

\begin{table}[h!]
\begin{center}
\caption{Inter-particle distances $r_{m1}^{(A)},\,r_{m2}^{(A)},...,\,r_{m8}^{(A)}$ (in
units $\sigma$) and the areas $S_{m1}^{(A)},\,S_{m2}^{(A)},...,\,S_{m5}^{(A)}$ (in
units $\sigma^{2}$), which correspond to positions of the main
maxima of the functions $g(r)$ and $g(S)$ in the case of amorphous
aluminium. The errors in the values are estimated to be less than
$8$\%.} \label{table_2}\centerline{}
\begin{tabular}[t]{ccccccccc}
\hline
$T,\,$K & $r_{m1}^{(A)}$ & $r_{m2}^{(A)}$ & $r_{m3}^{(A)}$ & $r_{m4}^{(A)}$ & $r_{m5}^{(A)}$ & $r_{m6}^{(A)}$ & $r_{m7}^{(A)}$ & $r_{m8}^{(A)}$ \\
\hline
$50$   & $1.0$ & $1.472$ & $1.731$ & $1.983$ & $2.381$ & $2.657$ & $2.962$  & $3.612$ \\
$100$  & $1.0$ & $1.465$ & $1.738$ & $1.982$ & $2.379$ & $2.654$ & $2.961$  & $3.609$ \\
$150$  & $1.0$ & $1.455$ & $1.734$ & $1.982$ & $2.378$ & $2.651$ & $2.958$  & $3.605$ \\
\hline\hline
$T,\,$K & $S_{m1}^{(A)}$ & $S_{m2}^{(A)}$ & $S_{m3}^{(A)}$ & $S_{m4}^{(A)}$ & $S_{m5}^{(A)}$ & & \\
\hline
$50$  & $0.02$ &  --  & $0.455$ & $0.804$ & $1.17$ & &\\
$100$  & $0.06$ & $0.24$ & $0.471$ & $0.811$ & $1.23$ & &\\
$150$  & $0.08$ & $0.28$ & $0.458$ & $0.851$ & $1.19$ & &\\
\hline
\end{tabular}
\end{center}
\end{table}

\begin{table}[h]
\begin{center}
\caption{Numerical values of the parameters $\upsilon_{1}$,
$\upsilon_{2}$, $\Omega_{1}^{2}$, and $p$ of the
integro-differential Eq.(\ref{eq_21}) obtained from fit of numerical
solution of Eq.(\ref{eq_21}) to simulation results.}
\label{table_3}\centerline{}
\begin{tabular}[t]{ccccc}
\hline
$T,\,$K & $\upsilon_{1}$ & $\upsilon_{2}$ & $\Omega_{1}^{2}$, $\times10^{25}\,$s$^{-2}$ & $p$ \\
\hline
$50$    & $1.531$ & $13.43$ & $3.3$  & $1.62$ \\
$100$   & $1.232$ & $9.334$ & $6.5$  & $1.52$ \\
$150$   & $0.934$ & $8.415$ & $9.8$  & $1.45$ \\
$1000$  & $0.815$ & $1.873$ & $64$   & $2.32$ \\
$1500$  & $0.795$ & $1.721$ & $98$   & $2.0$  \\
$2000$  & $0.713$ & $1.634$ & $130$  & $1.96$ \\
\hline
\end{tabular}
\end{center}
\end{table}

\end{document}